# Flare Impulsive-phase Durations

Brian R. Dennis,[1] Hugh Hudson,[2] and Joel Allred[3]

[1] *The Catholic University of America, Washington, DC*
Email: Brian.R.Dennis@nasa.gov

[2] *University of Glasgow*

[3] *NASA Goddard Space Flight Center, Solar Physics Laboratory, Code 671, Greenbelt, MD 20771*

## ABSTRACT

This Research Note is in response to the recent paper by S. M. Perriyil et al. (2026). They provide measurements of the time delay (Δt) between the hard X-ray and soft X-ray peak times for 96 flares observed with RHESSI and GOES. These delays are found to be dependent on the length of the magnetic loop(s) joining the HXR footpoints seen in RHESSI images. We offer a possible explanation for this coincidence in terms of the duration of the electron beam heating, commonly inferred from the duration of the HXR emission, and the time taken for heated plasma to rise to the loop top as inferred in this paper from Δt. We suggest that the particle acceleration occurs at or near the top of the loop(s) and that it is quenched by the increase in density as the heated plasma reaches the acceleration site.

*Keywords:* Solar physics (1476) — Solar flares (1496)) — Solar x-ray flares (1816)

RESPONSE

S. M. Perriyil et al. (2026) report the results of the analysis of 96 C, M, and X flares observed in soft X-rays with GOES and in hard X-rays with RHESSI, They found a correlation between the time delay (Δt) between the hard X-ray and soft X-ray peak times and the loop length (L) derived from the separation of the foot points imaged with RHESSI. They state that "…the delay between HXR and SXR peaks can be directly interpreted as the loop filling time required for evaporated plasma to reach its maximum thermal emission." They calculate the velocity ($V_{evap}$) of the up-flowing plasma associated with what is usually called "chromospheric evaporation" using the relation

$$V_{evap} = L/\Delta t \quad (1)$$

This description is not consistent with the strict interpretation of the standard Neupert Effect (e.g. A. Veronig et al. (2002)), in which the rise of the SXR emission is closely matched by the time integral of the HXR emission, with the SXR peaking at the end of the impulsive phase. This is commonly interpreted in a model in which flare-accelerated electrons travel down the legs of a magnetic loop and lose their energy by Coulomb collisions at the footpoints while producing the bremsstrahlung seen as the HXR emission. The electron collisional losses rapidly heat the chromospheric plasma and increase the pressure such that the heated plasma is driven up into the loop at high speed. Since the plasma cooling time is much longer than the heating time, the total amount of hot plasma in the loop increases as the time integral of the input energy and hence of the HXR emission. The amount of hot plasma in the loop stops rising and the SXR emission reaches its peak when the energy input from the accelerated electrons stops, corresponding to the end of the HXR emission. This results in the usually excellent agreement between the time derivative of the SXR emission and the HXR light curve. The SXR peak usually occurs when the HXR emission ends and there is no more plasma heating by the electron beam. A. Veronig et al. (2002) describe this in detail and conclude that "a large fraction of the events reveal a timing behavior that is consistent with the Neupert effect…" There are certainly events where this is not true but they are not the majority of the events that S. M. Perriyil et al. (2026) have analyzed. We define $T_{imp}$ to be the duration of the plasma heating by the flare-accelerated electrons as they deposit their energy at the loop footpoints. The time scales Δ$t$ and $T_{imp}$ need not be related physically.

S. M. Perriyil et al. (2026) do in fact find a correlation between L and Δt, and similar correlations have been reported earlier by P. Saint-Hilaire et al. (2008), S. Toriumi et al. (2017), and J. W. Reep & S. Toriumi (2017). Of course, it is possible that this relation is just because there is more available free energy in large active regions that produce long loops and it takes longer to release it. But why would the duration of the impulsive energy release closely match the time for the heated plasma to rise up to the top of the loop? One possible explanation for this relation is that the electron acceleration process in the corona continues until the heated plasma reaches the top of the loop, at which time it is quenched by the greatly increased density at that location where the acceleration is taking place. Hence, no more electrons are accelerated to >20 keV, no more HXRs are produced, and no more plasma heating takes place.

Of course, there are all sorts of caveats to this simple scenario. The SXR peak sometimes comes after the end of [49] the HXRs, suggesting additional heating. The HXR duration sometimes is very different as you go to higher energies. The SXR decay time is longer than the predicted plasma cooling time. But for the simplest events that do follow the standard Neupert Effect, this simple scenario is consistent with the observations and is useful as a baseline for understanding what might be going on in more complicated events.

REINTERPRETATION

We offer a possible reinterpretation of the three events giving the largest values of Δt in Figures 4, 5, and 6 of S. M. Perriyil et al. (2026). Figure 1 shows the RHESSI) and GOES light curves for their event (SOL2014-10-22T01:16) with the longest loop and the largest value of Δt. This event also has a lengthy $T_{imp}$. The relevant curves for determining Δt are the RHESSI 25 - 50 keV count rates shown in cyan and the GOES 1 - 8 Å irradiance (W/$m^2$) as shown in black. The HXR peak is at 14:06:16 UT and the SXR peak is at 14:28:16 UT giving Δt = 1320 s. Clearly, the HXR emission continues after the SXR peak in contradiction to the Neupert Effect. However, the HXR light curve (shown in cyan) shows three peaks, and these are matched by changes in the rate of rise of the SXR light curves. Furthermore, RHESSI images reveal HXR sources at different locations at the times of the three peaks suggesting that it should be broken down into three shorter times, one for each of the three episodes of energy release. The multiple HXR peaks in the 22 October 2014 event would then be explained as different acceleration episodes, possibly on different loops, each with its own hot plasma rising to the top of the loop to quench the acceleration process. This then would decrease Δt by at least a factor of two. Changing the values of Δt by similar amounts for all three of the highest points in their figures would change the mean plasma up-flow velocity determined from the least-squares fits to all the points to ~100 km/s, i.e. more consistent with values obtained from EUV line blue-shift measurements for similar events (E. Antonucci et al. 1982).

REFERENCES

Antonucci, E., Gabriel, A. H., Acton, L. W., et al. 1982, SoPh, 78, 107, doi: 10.1007/BF00151147

Perriyil, S. M., Sadangaya, S. S., Gim´enez de Castro, C. G., & Simoes, P. J. A. 2026, ApJ, 999, 27, doi: 10.3847/1538-4357/ae3061

Reep, J. W., & Toriumi, S. 2017, ApJ, 851, 4, doi: 10.3847/1538-4357/aa96fe

Saint-Hilaire, P., Krucker, S., & Lin, R. P. 2008, SoPh, 250, 53, doi: 10.1007/s11207-008-9193-9

Toriumi, S., Schrijver, C. J., Harra, L. K., Hudson, H., & Nagashima, K. 2017, ApJ, 834, 56, doi: 10.3847/1538-4357/834/1/56

Veronig, A., Vrˇsnak, B., Dennis, B. R., et al. 2002, A&A, 392, 699, doi: 10.1051/0004-6361:20020947

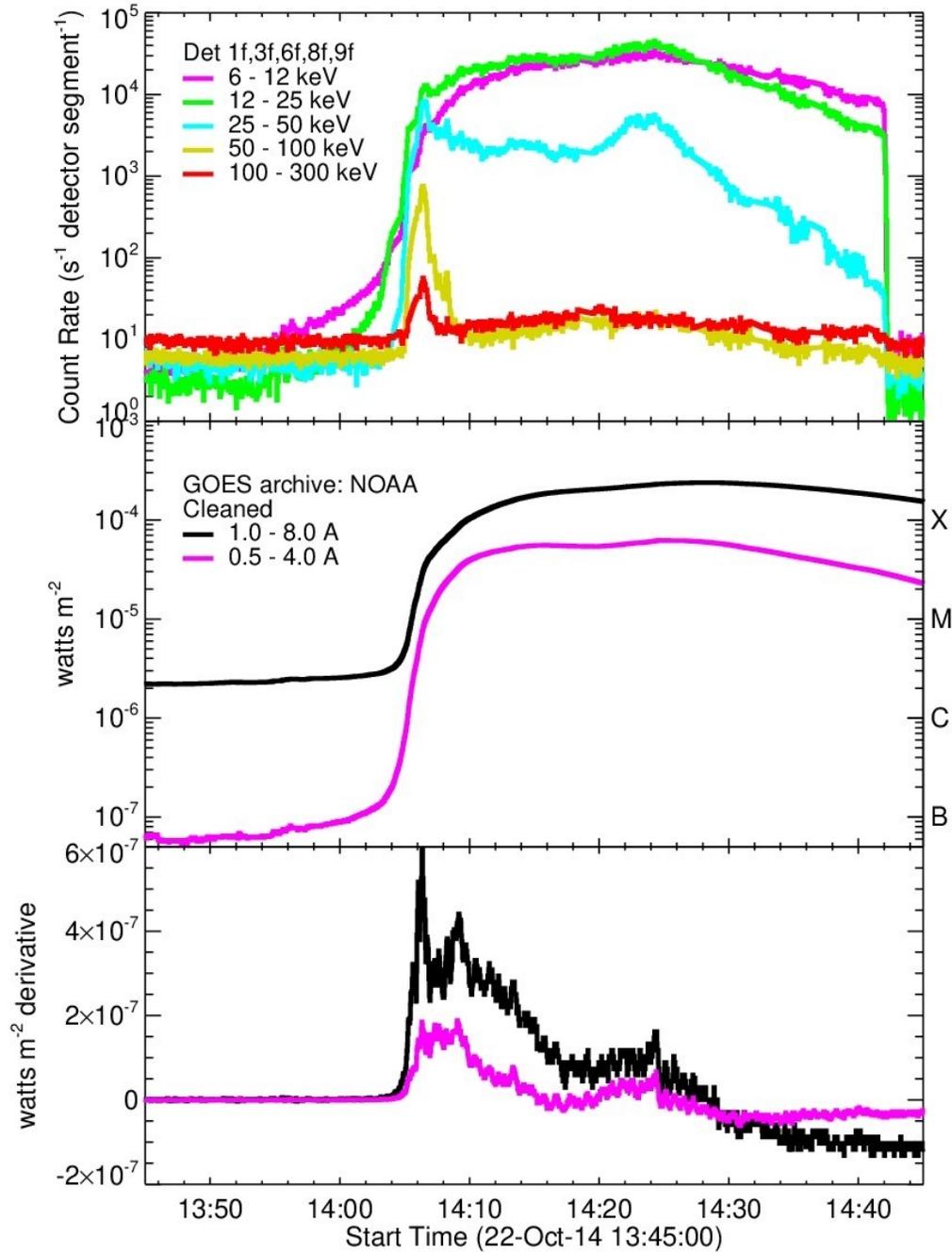

**Figure 1.** GOES SXR and RHESSI HXR light curves for SOL2014-10-22T14:02. The 25 - 50 keV light curve (cyan) shows three peaks suggestive of three separate episodes of particle acceleration possibly on different loops as indicated by the different locations of the sources imaged with RHESSI at these times. Thus, more realistic values for Δt might be the times between the HXR peaks, i.e. at least a factor 2 shorter than the value used by S. M. Perriyil et al. (2026) in their Figures 4, 5, and 6. This would then increase the mean velocity of the up-flowing heated plasma derived from the least-squares fits to the points in these figures to be closer to 100 km/s and similar to the range of velocities derived from EUV line blue shifts (E. Antonucci et al. 1982).